\documentclass{article}
\usepackage{amsmath,amsthm,amssymb}
\usepackage{array, longtable}
\newtheorem{Theorem}{Theorem}[section]
\newtheorem{lem}[Theorem]{Lemma}
\newtheorem{Remark}[Theorem]{Remark}
\newtheorem{Definition}[Theorem]{Definition}
\newtheorem{Corollary}[Theorem]{Corollary}
\newtheorem{Proposition}[Theorem]{Proposition}
\newtheorem{Example}[Theorem]{Example}

\numberwithin{equation}{section}

\begin{document}

\title{Application of  Constacyclic codes to Quantum MDS Codes\footnote{
 {\small Email addresses: bocong\_chen@yahoo.com (B. Chen),
  lingsan@ntu.edu.sg (S. Ling),
  zghui2012@126.com (G. Zhang).}}}

\author{Bocong Chen$^1$, ~San Ling$^1$, ~Guanghui Zhang$^2$}

\date{\small
${}^1$Division of Mathematical Sciences, School of Physical \& Mathematical Sciences,
         Nanyang Technological University, Singapore 637616, Singapore\\
${}^2$School of Mathematical Sciences,
Luoyang Normal University,
Luoyang, Henan, 471022, China}

\maketitle
\begin{abstract}
Quantum maximal-distance-separable (MDS) codes form an important class of quantum codes.
To get  $q$-ary quantum MDS codes,
it suffices to find linear MDS  codes $C$ over $\mathbb{F}_{q^2}$ satisfying $C^{\perp_H}\subseteq C$
by  the Hermitian construction and  the quantum Singleton bound.
If $C^{\perp_{H}}\subseteq C$, we say that $C$ is a  dual-containing code.
Many new quantum MDS codes with relatively
large minimum distance have been produced by constructing  dual-containing constacyclic  MDS codes (see \cite{Guardia11}, \cite{Kai13}, \cite{Kai14}).
These works motivate us to  make a careful study on the existence condition for nontrivial dual-containing constacyclic codes.
This would help us to avoid  unnecessary attempts and provide effective ideas in
order to construct dual-containing codes. Several classes of dual-containing MDS constacyclic codes are constructed and their parameters are computed. Consequently, new  quantum MDS codes are derived from these parameters. The quantum MDS codes  exhibited here have parameters better than the ones
available in the literature.


\end{abstract}

\section{Introduction}
Quantum  codes are useful in quantum computing and in quantum communications.
Just as in the classical case, any $q$-ary quantum code has three parameters, the code length,
the size of the code and the minimum  distance.
One of the principal problems in quantum error correction
is to construct quantum codes with the best possible minimum
distance. The  Hermitian construction, the CSS construction
and the symplectic
construction are the  frequently-used construction methods
(see, \cite{Laflamme}-\cite{Steane992},  \cite{Bierbrauer00}, \cite{Ashikhmin01}, \cite{Aly07}, \cite{Feng}-\cite{Hamada},  \cite{Jin10},  \cite{Jin12}).

Calderbank {\it et al.} in \cite{Calderbank} discovered that the  construction of quantum
error-correcting codes can be diverted into finding classical self-orthogonal
codes over $\mathbb{F}_2$ or $\mathbb{F}_4$ with respect to certain inner
product. After that, a lot of good  quantum codes were
obtained by using classical error-correcting codes (see \cite{Chau}, \cite{Cohen},\cite{hchen}, \cite{Jin10}, \cite{Kai13}).

We use $[[n,k,d]]_q$ to denote a $q$-ary quantum code of length $n$ with size $q^k$ and minimum distance $d$, where $q$
is a prime power. It is well known that  the
parameters of an  $[[n,k,d]]_q$ quantum code must
satisfy the quantum Singleton bound: $2d\leq n-k+2$ (see \cite{Knill} and \cite{Ketkar}).
A quantum code
beating this quantum Singleton bound is called a quantum
maximum-distance-separable (MDS) code.
The construction of quantum MDS codes has been extensively studied.
However, it is a challenge to construct quantum MDS codes
with length $n>q+1$. Moreover, constructing quantum MDS codes with relatively
large minimum distance turns out to be  difficult. As mentioned in \cite{Jin14}, except for some sparse lengths
$n$ such as $n=q^2+1, \frac{q^2+1}{2}$ and $q^2$,
almost all known $q$-ary quantum MDS codes have minimum distance less than
or equal to $\frac{q}{2}+1$.

Recently,
many new   quantum MDS codes have been obtained basing on
the following Hermitian construction (see \cite{Ashikhmin01}):

{\it Theorem 1.1 (Hermitian Construction):~ If $C$ is a $q^2$-ary $[n,k,d]$-linear code such that
$C^{\perp_H}\subseteq C$, then there exists a $q$-ary quantum code with parameters $[[n, 2k-n, \geq d]]_q$.}

To get  $q$-ary quantum MDS codes, the Hermitian construction and the quantum Singleton bound imply that
we only need to find linear MDS  codes $C$ over $\mathbb{F}_{q^2}$ satisfying $C^{\perp_H}\subseteq C$.
From this idea,  Guardia
in \cite{Guardia11} constructed a new class of quantum MDS codes based on cyclic codes;
Kai and Zhu in \cite{Kai13} obtained two new classes of  quantum MDS codes by using
negacyclic codes.
Following that line of research, Kai {\it et al.} in \cite{Kai14} produced several new quantum MDS codes based on
constacyclic codes. As pointed out in \cite{Kai14}, it turns out that
constacyclic codes are a good source of producing quantum MDS codes.

These works motivate us to  make a careful study on the condition $C^{\perp_H}\subseteq C$ when
$C$ is a constacyclic code.  If $C^{\perp_{H}}\subseteq C$, we say that $C$ is a  dual-containing code. We show that dual-containing $\lambda$-constacyclic codes  over $\mathbb{F}_{q^2}$  exist only when the order of $\lambda\in \mathbb{F}_{q^2}^*$ is a divisor of $q+1$.
Furthermore,
we obtain  elementary number-theoretic conditions for the existence of dual-containing constacyclic codes.
This would help us to avoid  unnecessary attempts and provide effective ideas in
order to construct dual-containing codes.

In this paper, several classes of dual-containing MDS constacyclic codes are constructed and their parameters are computed.
Consequently, new  quantum MDS codes are derived from these parameters.
More precisely, we construct four classes of $q$-ary
quantum MDS codes with parameters:
\begin{itemize}
\item[{(i)}]
$$[[\frac{q^2-1}{3},\frac{q^2-1}{3}-2d+2,d]]_q$$
where $q$ is an odd prime power with $3\mid (q+1)$ and $2\leq d\leq\frac{2(q-2)}{3}+1$.

\item[{(ii)}]
$$[[\frac{q^2-1}{5},\frac{q^2-1}{5}-2d+2,d]]_q$$
where $q$ is an odd prime power with $5\mid (q+1)$ and $2\leq d\leq\frac{3(q+1)}{5}-1$.

\item[{(iii)}]
$$[[\frac{q^2-1}{7},\frac{q^2-1}{7}-2d+2,d]]_q
$$
where $q$ is an odd prime power with $7\mid (q+1)$ and  $2\leq d\leq\frac{4(q+1)}{7}-1$.

\item[{(iv)}]
$$[[\frac{q^2+1}{10},\frac{q^2+1}{10}-2d+2,d]]_q$$
where  $q=10m+3$ or $q=10m+7$, and $3\leq d\leq4m+1$ is odd.
\end{itemize}
Comparing the parameters with  all known quantum MDS codes,  we find that
many of these quantum MDS codes are new in the sense that their parameters are not covered by the
codes available in the literature. More specifically, fixing the length and $q$, the new codes have minimum distance
greater than the ones available in the literature.
We also  mention that the quantum MDS codes given by $(i)$ has minimum distance $d>\frac{q}{2}+1$ when $q>11$.
When $q>19$ (resp. $q>27$),
the quantum MDS codes given by $(ii)$ (resp. $(iii)$) has minimum distance $d>\frac{q}{2}+1$.
\section{Existence conditions for nontrivial Hermitian dual-containing constacyclic codes}
In this section, we recall some definitions and basic properties of constacyclic codes (see  \cite{Dinh12} and \cite{Chen}),
and prove some results concerning the existence conditions for Hermitian dual-containing constacyclic codes.

\subsection{Review of constacyclic codes}
Let $\mathbb{F}_{q^2}$ be
the finite field with $q^2$ elements.
Let $\mathbb{F}_{q^2}^*$ denote the multiplicative group of  nonzero elements of $\mathbb{F}_{q^2}$.
For $\beta\in \mathbb{F}_{q^2}^*$ we denote by $\rm{ord}(\beta)$
the order of $\beta$ in the group $\mathbb{F}_{q^2}^*$;
then $\rm{ord}(\beta)$ is a divisor of $q^2-1$, and $\beta$
is called a {\em primitive $\rm{ord}(\beta)$th root of unity}.

Starting from this section till the end of this paper, we assume that $n$ is a positive integer relatively prime to $q$.
Let  $\mathbb{F}_{q^2}^n$  be the $\mathbb{F}_{q^2}$-vector space of $n$-tuples.
A {\it linear  code} $C$ of length $n$ over $\mathbb{F}_{q^2}$ is an $\mathbb{F}_{q^2}$-subspace of $\mathbb{F}_{q^2}^n$.
For $\lambda\in \mathbb{F}_{q^2}^*$, a linear code $C$ of length $n$ over $\mathbb{F}_{q^2}$ is said to be
{\it $\lambda$-constacyclic} if $(\lambda c_{n-1}, c_0,\cdots,c_{n-2})\in C$
for every $(c_{0}, c_1,\cdots,c_{n-1})\in C$.  When $\lambda$ =$1$,
$\lambda$-constacyclic codes are {\it cyclic codes}, and when $\lambda=-1$,
$\lambda$-constacyclic codes are just {\it negacyclic codes}.

Each codeword $\mathbf{c}=(c_0, c_1,\cdots, c_{n-1})\in C$ is customarily identified with its polynomial representation
$c(X)=c_0+c_1X+\cdots+c_{n-1}X^{n-1}$. In this way, every $\lambda$-constacyclic code $C$ is identified with exactly one ideal of
the quotient algebra $\mathbb{F}_{q^2}[X]/\langle X^n-\lambda\rangle$.
We  then know that $C$ is generated uniquely by a monic divisor $g(X)$ of $X^n-\lambda;$
in this case, $g(X)$ is called the {\it generator polynomial}
of $C$ and we write $C=\langle g(X)\rangle$.
In particular, the irreducible factorization of $X^n-\lambda$
in $\mathbb{F}_{q^2}[X]$ determines all $\lambda$-constacyclic
codes of length $n$ over $\mathbb{F}_{q^2}$.

Let  $\lambda\in \mathbb{F}_{q^2}^*$ be a primitive $r$th root of unity.
Then there exists a primitive $rn$th
root of unity (in some extension field of $\mathbb{F}_{q^2}$), say $\eta$, such that $\eta^n=\lambda$.
The roots of $X^n-\lambda$ are precisely the elements  $\eta^{1+ri}$ for $0\leq i\leq n-1$.
Set $\theta_{r,n}=\{1+ri\,|\,0\leq i\leq n-1\}$.
The defining set of a constacyclic code $C=\langle g(X)\rangle$ of length $n$ is the set
$Z=\{j\in \theta_{r,n}\,|\,\eta^j ~\hbox{is a root of} ~g(X)\}$.
It is easy to see that the defining set $Z$ is  a union of some $q^2$-cyclotomic cosets modulo $rn$ and
$\dim_{\mathbb{F}_{q^2}}(C)=n-|Z|$ (see \cite{Yang}).

The following results are  important  in constructing
quantum codes. (see \cite[Theorem 2.2]{Aydin} or \cite[Theorem 4.1]{Yang}).
\begin{Theorem}(The BCH bound for Constacyclic Codes)\label{BCH}
Let $C$ be a $\lambda$-constacyclic code of length $n$ over $\mathbb{F}_{q^2}$, where $\lambda$ is a primitive $r$th root of unity.
Let $\eta$ be  a primitive $rn$th root of unity in an extension field of $\mathbb{F}_{q^2}$ such that $\eta^n=\lambda$.
Assume the generator polynomial of $C$ has roots that include the set $\{\eta\zeta^i\,|\, i_1\leq i\leq i_1+d-1\}$, where
$\zeta=\eta^r$. Then the minimum distance of $C$ is at least $d$.
\end{Theorem}

\begin{Proposition}
(Singleton  bound for linear codes)  If an  $[n,k,d]$ linear code over $\mathbb{F}_{q^2}$ exists, then $k\leq n-d+1$.
\end{Proposition}

The {\it Hermitian inner product}
is defined as
$$
\mathbf{x}\ast\mathbf{y}=x_0\overline{y_0}+x_1\overline{y_1}+\cdots+x_{n-1}\overline{y_{n-1}},
$$
where $\mathbf{x}=(x_0,x_1,\cdots,x_{n-1})\in \mathbb{F}_{q^2}^n$, $\mathbf{y}=(y_0,y_1,\cdots,y_{n-1})\in \mathbb{F}_{q^2}^n$
and $\overline{y_i}=y_i^{q}$.

The {\it Hermitian dual code} of $C$ is defined as
$$
C^{\perp_{H}}=\Big\{\mathbf{x}\in \mathbb{F}_{q^2}^n\,|\,\sum\limits_{i=0}^{n-1}x_i\overline {y_i}=0, \forall~\mathbf{y}\in C\Big\}.
$$
If $C\subseteq C^{\perp_{H}}$, then $C$ is called a
(Hermitian)   self-orthogonal code.
Conversely, if $C^{\perp_{H}}\subseteq C$, we say that $C$ is a (Hermitian) dual-containing code.
Dual-containing codes are also known as weakly self-dual codes;
it is the type of code we are most concerned with in this
paper.
Clearly, $\{0\}$ is a self-orthogonal code and $\mathbb{F}_{q^2}^n$ is a dual-containing code.
These two codes are called {\it trivial codes}.

On the other hand,
observe that the automorphism of $\mathbb{F}_{q^2}$ given by  $``^{-}"$,  $^{-}(x)=\overline{x}=x^{q}$ for any $x\in \mathbb{F}_{q^2}$, can be
extended to an automorphism of $\mathbb{F}_{q^2}[X]$ in an obvious way:
$$
~~\mathbb{F}_{q^2}[X]~\longrightarrow ~\mathbb{F}_{q^2}[X],~~~\sum\limits_{i=0}^na_iX^i~~\mapsto~~\sum\limits_{i=0}^n\overline{a_i}X^i,~~~
\hbox{for any $a_0,a_1, \cdots, a_{n}$ in $\mathbb{F}_{q^2}$},
$$
which is also denoted by $``^-"$ for simplicity.

For a monic polynomial $f(X)\in \mathbb{F}_{q^2}[X]$ of degree $k$ with $f(0)\neq0$, its {\it reciprocal polynomial} $f(0)^{-1}X^kf(X^{-1})$ will be denoted by $f(X)^*$. Note that $f(X)^*$ is also a monic polynomial.

The following result gives the generator polynomial of $C^{\perp_{H}}$.
\begin{lem}(\cite[Lemma 2.1(ii)]{Yang})\label{first}
Let $C=\langle g(X)\rangle$ be a $\lambda$-constacyclic code of length $n$ over $\mathbb{F}_{q^2}$,
where $g(X)$ is the generator polynomial of $C$. Let $h(X)=\frac{X^n-\lambda}{g(X)}$.
Then the Hermitian dual code $C^{\perp_{H}}$ is an $\overline{\lambda}^{-1}$-constacyclic code with generator polynomial
$\overline{h(X)^*}$.
\end{lem}

\begin{Remark}\label{sigma}
Let $f(X)$ be a monic polynomial in $\mathbb{F}_{q^2}[X]$ with $f(0)\neq0$.
It is readily seen that $\overline{f(X)^*}=\big(\overline{f(X)}\big)^*$.
For simplicity we write $f(X)^{\sigma}=\overline{f(X)^*}=\overline{f(X)}^*$,
namely $\sigma$ can be regarded as the composition $``^{-}\circ^{*}"$.
It is clear that $f(X)^{\sigma^2}=f(X)$.
\end{Remark}

Motivated by Theorem $1.1$, we aim  to determine existence conditions for nontrivial
dual-containing constacyclic codes.
We first need the following observation.
\begin{lem}\label{reduction}
Let
$\alpha, \beta$ be nonzero elements of $\mathbb{F}_{q^2}$.
Let $C_1, C_2$ be nontrivial (i.e., neither $\{0\}$ nor $\mathbb{F}_{q^2}^n$) $\alpha$-and $\beta$-constacyclic codes of length $n$ over $\mathbb{F}_{q^2}$, respectively.
If $C_1\subseteq C_2$,  then $\alpha=\beta$.
\end{lem}
\begin{proof}
The proof is nearly identical to the proof of Proposition 3.1 in \cite{Dinh13}.
Suppose otherwise that $\alpha\neq\beta$. As $C_1$ is nonzero,
without loss of generality, we can assume that $(c_0, c_1, \cdots,  c_{n-1})\in C_1$ with $c_{n-1}\neq0$.
It follows that both $(\alpha c_{n-1}, c_0, \cdots,  c_{n-2})$ and $(\beta c_{n-1}, c_0, \cdots,  c_{n-2})$
belong to $C_2$. This implies that
$$
(1, 0, \cdots, 0)=(\alpha-\beta)^{-1}c_{n-1}^{-1}\big((\alpha c_{n-1}, c_0, \cdots,  c_{n-2})-(\beta c_{n-1}, c_0, \cdots,  c_{n-2})\big)\in C_2.
$$
We therefore deduce that $(1, 0, \cdots, 0), (0, 1, \cdots, 0), \cdots, (0, 0, \cdots, 1)$ belong to $C_2$. This is a contradiction.
\end{proof}

As an immediate application of Lemma~\ref{first} and  Lemma~\ref{reduction}, we have the following result.
\begin{Corollary}\label{restrict}
Let $\lambda\in\mathbb{F}_{q^2}^*$ be a primitive $r$th root of unity and let
$C$ be a nontrivial Hermitian dual-containing  $\lambda$-constacyclic code of length $n$ over $\mathbb{F}_{q^2}$.
We then have $\lambda=\overline{\lambda}^{-1}$, i.e., $r\mid(q+1)$.
\end{Corollary}

The next result presents  a criterion to determine whether or not a given  $\lambda$-constacyclic code of
length $n$ over $\mathbb{F}_{q^2}$ is dual-containing.
\begin{lem}(see \cite[Lemma 2.2]{Kai14})
Let $r$ be a positive divisor of $q+1$ and let $\lambda\in \mathbb{F}_{q^2}^*$ be of order $r$. Assume that $C$ is a $\lambda$-constacyclic
code of length $n$ over $\mathbb{F}_{q^2}$ with defining set $Z$. Then $C$ is a dual-containing code if and only if $Z\bigcap(-qZ)=\emptyset$,
where $-qZ=\{-qz (\bmod ~rn)\,|\,z\in Z\}$.
\end{lem}

\subsection{Existence conditions for nontrivial dual-containing constacyclic codes}
Assume that $\lambda\in \mathbb{F}_{q^2}$ is a primitive $r$th root of unity. Clearly, $r$ is a divisor of $q^2-1$.
In particular, $\gcd(r,q)=1$. To study nontrivial dual-containing
$\lambda$-constacyclic codes, we may assume first that  $\lambda=\overline{\lambda}^{-1}$ by  Corollary~\ref{restrict},
i.e.,  $r\mid(q+1)$.

For any monic irreducible factor
$f(X)\in \mathbb{F}_{q^2}[X]$ of $X^{n}-\lambda$,
$f(X)^{\sigma}$ is also a monic irreducible factor of $X^{n}-\lambda$ satisfying $f(X)^{\sigma^2}=f(X)$ (see Remark~\ref{sigma}).
This implies that $X^{n}-\lambda$ can be factorized into distinct monic irreducible polynomials
as follows
$$
X^{n}-\lambda=f_1(X)f_2(X)\cdots f_u(X)h_1(X)h_1^{\sigma}(X)h_2(X)h_2^{\sigma}(X)\cdots h_v(X)h_v^{\sigma}(X)
$$
where $f_i(X)$ ($1\leq i\leq u$) are distinct monic irreducible factors over $\mathbb{F}_{q^2}$ such that $f_i(X)^{\sigma}=f_i(X)$
while $h_{j}(X)$ and $h_j(X)^{\sigma}$ ($1\leq j\leq v$) are distinct monic irreducible factors over $\mathbb{F}_{q^2}$.
 As such, we have the following definition:
\begin{Definition}
Let $f(X)$ be a monic polynomial in $\mathbb{F}_{q^2}[X]$ with $f(0)\neq0$.
We say that $f(X)$ is conjugate-self-reciprocal if $f(X)^{\sigma}=f(X)$. Otherwise,
we say that $f(X)$ and $f(X)^{\sigma}$ form a conjugate-reciprocal polynomial
pair.
\end{Definition}

It should be pointed out that $u$ may be equal to $0$, namely any irreducible factor of $X^{n}-\lambda$ over $\mathbb{F}_{q^2}$ is not conjugate-self-reciprocal.
Likewise, it is possible that $v=0$, namely every irreducible factor of $X^{n}-\lambda$ over $\mathbb{F}_{q^2}$ is  conjugate-self-reciprocal.

Let $C=\langle g(X)\rangle$ be a nontrivial $\lambda$-constacyclic code of length $n$ over $\mathbb{F}_{q^2}$,
where $g(X)$ is a monic divisor of $X^n-\lambda$. We may assume, therefore, that
$$
g(X)=f_1(X)^{a_1}\cdots f_u(X)^{a_u}h_1(X)^{b_1}(h_1^{\sigma}(X))^{c_1}\cdots h_v(X)^{b_v}(h_v^{\sigma}(X))^{c_v}
$$
where $0\leq a_i\leq 1$ for each $i$, and $0\leq b_j, c_j\leq 1$ for each $j$.
Then the generator polynomial of $C^{\perp_{H}}$ is
$$
h(X)^{\sigma}=\overline{h(X)^*}=f_1(X)^{1-a_1}\cdots f_u(X)^{1-a_u}h_1(X)^{1-c_1}(h_1^{\sigma}(X))^{1-b_1}\cdots h_v(X)^{1-c_v}(h_v^{\sigma}(X))^{1-b_v}.
$$
By Lemma~\ref{first}, $C$ satisfies $C^{\perp_{H}}\subseteq C$ if and only if $g(X) \mid h(X)^{\sigma}$,
i.e.,
\begin{equation}\label{self-orthogonal}
\left\{
                   \begin{array}{ll}
                 2a_i \leq 1,& \hbox{for each $i$}, \\
b_j+c_j \leq 1, & \hbox{for each $j$}.
                   \end{array}
                 \right.
\end{equation}
It follows that $C=\langle g(X)\rangle$ satisfies $C^{\perp_{H}}\subseteq C$ if and only if
$$
C=\langle h_1(X)^{b_1}(h_1^{\sigma}(X))^{c_1}\cdots h_v(X)^{b_v}(h_v^{\sigma}(X))^{c_v}\rangle
$$
where $0\leq b_j, c_j\leq 1$ and $b_j+c_j \leq 1$  for each $j$. This discussion leads to the following result.

\begin{Theorem}\label{polynomial-orthogonal}
Let $\lambda\in \mathbb{F}_{q^2}^*$ satisfy $\lambda=\overline{\lambda}^{-1}$.
Nontrivial Hermitian dual-containring $\lambda$-constacyclic codes of
length $n$ over $\mathbb{F}_{q^2}$ exist if and only if $v>0$, i.e., there  exists at least one
conjugate-reciprocal polynomial
pair among the  monic irreducible factors
of $X^{n}-\lambda$ over $\mathbb{F}_{q^2}$.
\end{Theorem}

In the rest of this subsection we aim to  obtain more simplified criteria for the existence of nontrivial
Hermitian dual-containing $\lambda$-constacyclic codes of length $n$ over $\mathbb{F}_{q^2}$.

It is well known that the irreducible factors of $X^{rn}-1$ over $\mathbb{F}_{q^2}$ can be described via the $q^2$-cyclotomic cosets modulo $rn$
(see  \cite[Theorem 4.1.1]{Huffman}):
Assume that $\Omega=\{i_0=0, i_1=1, i_2, \cdots, i_{\rho}\}$
is a set of representatives of the $q^2$-cyclotomic cosets modulo $rn$;
let $C_{i_j}$ be the $q^2$-cyclotomic coset modulo $rn$ containing $i_j$
for $0\leq j\leq\rho$;
we then know that
\begin{equation}\label{simple-irr-decomposition}
X^{rn}-1= M_{{i_0}}(X)M_{{i_1}}(X)
 \cdots M_{{i_\rho}}(X)
\end{equation}
with
$$M_{{i_j}}(X)=\prod\limits_{s\in C_{i_j}}(X-\eta^s),
 \qquad j=0,\cdots,\rho,$$
all being monic irreducible in $\mathbb{F}_{q^2}[X]$, where $\eta$ is a primitive $rn$th root of unity over some extension field of $\mathbb{F}_{q^2}$
such that $\eta^n=\lambda$.
Since $X^{n}-\lambda$ is a  divisor of  $X^{rn}-1$ in $\mathbb{F}_{q^2}[X]$,
we can find a subset  $\Delta$ of $\Omega$  such that
\begin{equation}\label{co-prime-case}
X^{n}-\lambda=\prod\limits_{e\in\Delta}M_{e}(X).
\end{equation}
Set $\mathcal{O}_{r,n}=\{C_j\,|\,j\in\Delta\}$. We also see that $C_{i_1}=C_1\in \mathcal{O}_{r,n}$.
With these preparations, we  translate Theorem~\ref{polynomial-orthogonal}  into the language of
$q^2$-cyclotomic cosets modulo $rn$.

\begin{lem}\label{lem-orthogonal}
Let $\lambda\in \mathbb{F}_{q^2}^*$ be of order $r$ satisfying $\lambda=\overline{\lambda}^{-1}$.
There  exists a nontrivial dual-containing $\lambda$-constacyclic code of length $n$ over $\mathbb{F}_{q^2}$
if and only if there exists   $C_{e_0}\in\mathcal{O}_{r,n}$ such that $C_{e_0}\neq C_{-qe_0}$, where $C_{e_0}$ and $C_{-qe_{0}}$
denote the $q^2$-cyclotomic cosets modulo $rn$ containing $e_0$ and $-qe_0$, respectively.
\end{lem}
\begin{proof}
Let $M_{j}(X)=\prod_{i\in C_j}(X-\eta^i)$ be the minimal polynomial of $\eta^j$ over $\mathbb{F}_{q^2}$.
Note that $M_{j}(X)^*=M_{-j}(X)$.
Combining Theorem~\ref{polynomial-orthogonal} with (\ref{co-prime-case}), it suffices to prove that
$\overline{M_j(X)}=M_{qj}(X)$. For this purpose, we only need to show that $\eta^{qj}$ is a root of
$\overline{M_j(X)}$.
Assume that $M_j(X)=a_0+a_1X+\cdots+a_tX^t$ with $a_0, a_1,\cdots, a_t\in \mathbb{F}_{q^2}$.
Thus $\overline{M_j(X)}=\overline{a_0}+\overline{a_1}X+\cdots+\overline{a_t}X^t$.
Obviously $\overline{M_j(\eta^{qj})}=0$, since
$$
\overline{M_j(\eta^{qj})}=\overline{a_0}+\overline{a_1}\eta^{qj}+\cdots+\overline{a_t}(\eta^{qj})^t
=\big(a_0+a_1\eta^{j}+\cdots+a_t\eta^{tj}\big)^{q}=M_j(\eta^j)^q=0.
$$
\end{proof}
Let $\lambda\in \mathbb{F}_{q^2}^*$ be of order $r$ satisfying $\lambda=\overline{\lambda}^{-1}$.
Assume that $b$ is the highest power of $2$ dividing $r$, say $2^b\Vert r$, namely $2^b\mid r$ but $2^{b+1}\nmid r$.
Write $rn=2^{a+b}p_1^{k_1}p_2^{k_2}\cdots p_s^{k_s}$,
where $p_j$ are distinct odd primes and $k_j$ are positive integers for $1\leq j\leq s$.
Let $\mathbb{Z}_{m}^*$ denote the multiplicative group of all
residue classes modulo $m$ which are coprime with $m$.
Let ${\rm ord}_{p_j^{k_j}}(q)=2^{x_j}y_j$ such that $\gcd(2,y_j)=1$ for  $1\leq j\leq s$, where ${\rm ord}_{p_j^{k_j}}(q)$ denotes the multiplicative
order of $q\in \mathbb{Z}_{p_j^{k_j}}^*$.
We assert that  $2^{x_j}\Vert{\rm ord}_{p_j}(q)$ if and only if $2^{x_j}\Vert{\rm ord}_{p_j^{k_j}}(q)$ for  $1\leq j\leq s$.
Indeed, consider the natural surjective homomorphism
$\pi:~\mathbb{Z}_{p_j^{k_j}}^*~\mapsto~\mathbb{Z}_{p_j}^*$, $x~(\bmod~p_j^{k_j})~\mapsto~x~(\bmod~p_j)$.
We then know that ${\rm ord}_{p_j}(q)$
is exactly equal to the order of $qKer\pi$ in the factor group $\mathbb{Z}_{p_j^{k_j}}^*/Ker\pi$, which is also equal to the smallest positive integer $k$ such that $q^k\in Ker\pi$. Now the desired result follows from the fact that $Ker\pi$ is a group of odd order.

The next two results give existence conditions for nontrivial Hermitian dual-containing $\lambda$-constacyclic codes according to
the scopes of $a+b$.
\begin{Theorem}\label{dual-containing1}
With respect to the above notation, we assume further that
$a+b\leq1$.
Nontrivial Hermitian dual-containing $\lambda$-constacyclic codes of length $n$ over $\mathbb{F}_{q^2}$ exist
if and only if  one of the following statements holds:
\begin{itemize}
\item[{\bf(i)}]There exists an integer $t$, $1\leq t\leq s$,  such that $x_t=0$.

\item[{\bf(ii)}]The integers $x_j$ are greater than $1$ for all $1\leq j\leq s$ and $x_1=x_2=\cdots=x_s$.

\item[{\bf(iii)}]The integer $s\geq2$,
 $x_j>0$ for all $1\leq j\leq s$,  and
there exist distinct integers $j_{1}, j_{2}$ with $1\leq j_{1}, j_{2}\leq s$ such that
$x_{j_{1}}\neq x_{j_{2}}$.
\end{itemize}
\end{Theorem}
\begin{proof}
Suppose  that one of the above three conditions holds true, we work by contradiction to show that  nontrivial
Hermitian dual-containing $\lambda$-constacyclic codes of length $n$ over $\mathbb{F}_{q^2}$ exist.
By Lemma~\ref{lem-orthogonal}, we can suppose  that $C_{e}=C_{-qe}$ for any $C_e\in\mathcal{O}_{r,n}$, where $C_e$
denotes the $q^2$-cyclotomic coset modulo $rn$ containing $e$.
This leads to  $C_{1}=C_{-q}$ since $C_1\in\mathcal{O}_{r,n}$, which implies that an integer
$i_0'$  can be found such that $q^{1+2i_0'}\equiv-1~(\bmod~rn)$.
Let $i_0=2i_0'+1$, and thus $q^{i_0}\equiv-1~(\bmod~rn)$. Clearly,  $i_0$ is odd.

Assume that ${\rm (i)}$ holds.
There is no loss of generality to assume that $x_1=0$.
It follows from $q^{i_0}\equiv-1~(\bmod~p_1)$ that
$q^{2i_0}\equiv1~(\bmod~p_1)$.  Hence ${\rm ord}_{p_1}(q)\mid i_0$ as ${\rm ord}_{p_1}(q)$
is odd. This leads to $q^{i_0}\equiv1~(\bmod~p_1)$,
a contradiction.

Assume that ${\rm (ii)}$ holds. In particular, $x_1\geq2$.  Recall that ${\rm ord}_{p_1^{k_1}}(q)=2^{x_1}y_1$ such that $\gcd(2,y_1)=1$.
From $q^{i_0}\equiv-1~(\bmod~p_1^{k_1})$, we deduce that
$q^{2i_0}\equiv1~(\bmod~p_1^{k_1})$.  Hence, $2^{x_1}y_1$ divides $2i_0$, which  implies that $i_0$ is even.
This is a contradiction.

Now we assume that   ${\rm (iii)}$ holds.
Without loss of generality, we may assume that $x_1>x_2\geq1$.
From $q^{i_0}\equiv-1~(\bmod~p_j^{k_j})$ for all
$1\leq j\leq s$, we have  $q^{2i_0}\equiv1~(\bmod~p_j^{k_j})$.
Thus,
$2^{x_j}y_j\mid 2i_0$ and so
$2^{x_j-1}y_j\mid i_0$ for all $1\leq j\leq s$. In particular,
$2^{x_1-1}y_1\mid i_0$. Combining the fact $2^{x_1-1}y_1\mid i_0$ with the hypothesis $x_1>x_2\geq 1$,
it follows that $i_0$ is even,  a contradiction again.

Conversely, assume that  nontrivial Hermitian dual-containing
$\lambda$-constacyclic codes of length $n$ over $\mathbb{F}_{q^2}$ exist.  We assume
further that  neither {\rm {(i)}} nor {\rm {(iii)}} holds.
Then $x_j\geq1$ for all  $1\leq j\leq s$.  If $s=1$, we need to show that $x_1>1$. If $s\geq2$, we know that $x_1=x_2=\cdots=x_s>0$.
We are thus left to prove that  $x_1=x_2=\cdots=x_s=x>1$.
Suppose otherwise that $x=1$. Thus, $q^{2y_j}\equiv1~(\bmod~p_j^{k_j})$ for any $j$.
From the fact that $\mathbb{Z}_{p_j^{k_j}}^*$ is a cyclic group
whose unique element of order 2 is $[-1]_{p_j^{k_j}}$, where $[-1]_{p_j^{k_j}}$ denotes the residue class $\bmod~p_j^{k_j}$ containing $-1$.
It follows that $q^{y_j}\equiv-1~(\bmod~p_j^{k_j})$.
Let $y=\prod_{j=1}^sy_j$.
We get $q^{y}\equiv-1~(\bmod~p_j^{k_j})$ for all $1\leq j\leq s$.
Therefore, $q^{y}\equiv-1~(\bmod~p_1^{k_1}p_2^{k_2}\cdots p_s^{k_s})$.
This leads to  $q^{y}\equiv-1~(\bmod~rn)$, as $a+b\leq1$.
We get the desired contradiction, since we would obtain $C_1=C_{-q}$.

\end{proof}

Finally we consider the remainding case: $a+b\geq2$.
\begin{Theorem}\label{dual-containing2}
With respect to the above notation, we assume further that
$a+b\geq2$.
Nontrivial Hermitian dual-containing $\lambda$-constacyclic codes of length $n$ over $\mathbb{F}_{q^2}$ exist
if and only if  one of the following statements holds:
\begin{itemize}
\item[{\bf(i)}]$q\equiv1~(\bmod~4)$.
\item[{\bf(ii)}]$q\equiv-1~(\bmod~4)$ and $a+b>e$,
where $e$ is the  positive
integer such that  $2^e\Vert(q+1)$.
\item[{\bf(iii)}]There exists an integer $j$,  $1\leq j\leq s$, such that $x_{j}=0$.
\item[{\bf(iv)}]$x_j$ is nonzero for all $1\leq j\leq s$ and there  exists some integer $j_1$,
$1\leq j_1\leq s$, such that  $x_{j_1}\geq2$.
\end{itemize}
\end{Theorem}
\begin{proof}
By Lemma~\ref{lem-orthogonal}, we know that nontrivial Hermitian
dual-containing $\lambda$-constacyclic codes of length $n$ over $\mathbb{F}_{q^2}$ do not exist
if and only if $C_{1}=C_{-q}$,  where $C_{1}$ and $C_{-q}$
denote the $q^2$-cyclotomic cosets modulo $rn$ containing $1$ and $-q$, respectively.

Suppose that one of the above four conditions holds true, and we proceed by way of contradiction.
It follows from $C_1=C_{-q}$ that an odd integer
$i_0$  can be found such that $q^{i_0}\equiv-1~(\bmod~rn)$.

Assume that {\rm (i)} holds.
We have $q^{i_0}\equiv-1~(\bmod~2^{a+b})$,
since $2^{a+b}$ divides $rn$. By assumption $a+b\geq2$, so $q^{i_0}\equiv-1~(\bmod~4)$.
This contradicts $q\equiv1~(\bmod~4)$.

Assume that {\rm (ii)} holds.
Write $q+1=2^ef$, where $f$ is an odd positive integer.
By assumption $a+b>e$, then  $q^{i_0}\equiv-1~(\bmod~2^{e+1})$.
 Let $i_0=2i_0'+1$.
Since $q\equiv-1~(\bmod~2^{e})$, it follows that  $q^2\equiv1~(\bmod~2^{e+1})$, which gives $q^{2i_0'}\equiv1~(\bmod~2^{e+1})$.
Thus $q^{2i_0'+1}\equiv q~(\bmod~2^{e+1})$, namely $q^{i_0}\equiv q~(\bmod~2^{e+1})$.
Combining with $q^{i_0}\equiv -1~(\bmod~2^{e+1})$, we get $q\equiv -1~(\bmod~2^{e+1})$.
However,  this contradicts the fact that $q+1=2^ef$ with $f$ odd.

Assume that {\rm (iii)} holds. There is no loss of generality to assume that ${\rm ord}_{p_1}(q)$ is odd.
From $q^{i_0}\equiv-1~(\bmod~rn)$, we see that $q^{i_0}\equiv-1~(\bmod~p_1)$
and so $q^{2i_0}\equiv1~(\bmod~p_1)$. Since ${\rm ord}_{p_1}(q)\,|\,2i_0$, we have ${\rm ord}_{p_1}(q)\,|\,i_0$.
Thus $q^{i_0}\equiv1~(\bmod~p_1)$,
a contradiction.

Assume that {\rm (iv)} holds.
Recall that  ${\rm ord}_{p_j^{k_j}}(q)=2^{x_j}y_j$, where $\gcd(2,y_j)=1$ for each $1\leq j\leq s$.
Recall also that $2^{x_j}\Vert{\rm ord}_{p_j}(q)$ if and only if $2^{x_j}\Vert{\rm ord}_{p_j^{k_j}}(q)$.
Suppose $x_{1}\geq2$. Obviously $q^{2i_0}\equiv1~(\bmod~p_{1}^{k_1})$.
From ${\rm ord}_{p_{1}^{k_{1}}}(q)=2^{x_1}y_1$, it follows that $2^{x_1}y_1\mid2i_0$ and then $i_0$ is even.
This is  a contradiction.

Now,
suppose that
nontrivial Hermitian dual-containing $\lambda$-constacyclic codes of length $n$ over $\mathbb{F}_{q^2}$ exist.
Assume further that
{\rm (i)}, {\rm (ii)} and {\rm (iii)} do not hold.
We need to show that {\rm (iv)} holds.
Since {\rm (iii)} does not hold, ${\rm ord}_{p_j}(q)$ is even for all $j$.
Assume, by way of contradiction, that
${\rm ord}_{p_j}(q)$ is even but not divisible by $4$  for all $1\leq j\leq s$,
i.e.,
$x_j=1$ for all $1\leq j\leq s$.
It follows from $q^{2y_j}\equiv1~(\bmod~p_j^{k_j})$ that $q^{y_j}\equiv-1~(\bmod~p_j^{k_j})$.
Let $y=\prod_{j=1}^sy_j$.
We get $q^{y}\equiv-1~(\bmod~p_j^{k_j})$ for all $1\leq j\leq s$.
Therefore, $q^{y}\equiv-1~(\bmod~p_1^{k_1}p_2^{k_2}\cdots p_s^{k_s})$.
The assumption that neither {\rm (i)} nor {\rm (ii)} holds true implies that $2^{a+b}\mid(q+1)$.
It follows that  $q^{y}\equiv-1~(\bmod~2^{a+b})$, since $y$ is an odd positive integer.
Hence $q^{y}\equiv-1~(\bmod~rn)$.  This gives the desired contradiction.
\end{proof}

\begin{Example}
Let $q=11$, then $q^2=11^2$. suppose  $\mathbb{F}^*_{11^2}=\langle \theta\rangle$.
Let $\lambda=\theta^{10}$, then $r=12$. By Theorem~\ref{dual-containing2},
nontrivial Hermitian dual-containing $\lambda$-constacyclic codes of length $27$ over $\mathbb{F}_{121}$ do not exist.
This is because $rn=324=2^2\cdot 3^4$,   $a+b\geq 2$ and $q=11\equiv -1~(\bmod~4)$,
but $a+b = e=2$, and ${\rm ord}_3(11)=2$.

\end{Example}

\begin{Example}
Let $q=3^2$, then $q^2=3^4$. suppose  $\mathbb{F}^*_{3^4}=\langle \theta\rangle$.
Let $\lambda=\theta^8$, then $r=10$.

$(1)$ By Theorem~\ref{dual-containing1},
nontrivial  Hermitian dual-containing $\lambda$-constacyclic codes of length $5$ over $\mathbb{F}_{3^4}$ do not exist.
This is because  $rn=50=2\cdot 5^2$,  $a+b\leq 1$ and ${\rm ord}_5(9)=2$.

$(2)$  By Theorem~\ref{dual-containing2},  nontrivial Hermitian dual-containing $\lambda$-constacyclic codes of length $10$ over $\mathbb{F}_{3^4}$ exist.
Since $rn=100=2^2\cdot5^2$, then $a+b\geq 2$ and $q=9\equiv1~(\bmod~4)$.

\end{Example}

Applying Theorem~\ref{dual-containing1} and Theorem~\ref{dual-containing2} we have the following results, which will be used later for construction
quantum MDS codes.
\begin{Corollary}\label{exist1}
Let $q$ be an odd prime power.
Let $r$ be a positive integer dividing $q+1$ and let $n=\frac{q^2-1}{r}$. Assume that  $\lambda\in \mathbb{F}_{q^2}^*$ is of order $r$.
Then nontrivial Hermitian dual-containing $\lambda$-constacyclic codes of length $n$ over $\mathbb{F}_{q^2}$ exist.
\end{Corollary}
\begin{proof}
Clearly, $rn=q^2-1$, and so $4\mid rn$. If $q\equiv1~(\bmod~4)$, then we know from Theorem~\ref{dual-containing2}(i) that
the desired result follows. Otherwise, $q\equiv-1~(\bmod~4)$. In this case, the conditions satisfy Theorem~\ref{dual-containing2}(ii).
\end{proof}

\begin{Corollary}\label{exist2}
Let $q$ be an odd prime power such that $10\mid(q^2+1)$.
Let $r=q+1$ and  $n=\frac{q^2+1}{10}$. Assume that  $\lambda\in \mathbb{F}_{q^2}^*$ is of order $r$.
Then nontrivial Hermitian dual-containing $\lambda$-constacyclic codes of length $n$ over $\mathbb{F}_{q^2}$ exist.
\end{Corollary}
\begin{proof}
Let $rn=2^{a+b}p_1^{k_1}p_2^{k_2}\cdots p_s^{k_s}$,
where $p_j$ are distinct odd primes and $k_j$ are positive integers for $1\leq j\leq s$.
Let ${\rm ord}_{p_j^{k_j}}(q)=2^{x_j}y_j$ such that $\gcd(2,y_j)=1$ for  $1\leq j\leq s$.
We then know that ${\rm ord}_{rn}(q)={\rm lcm}({\rm ord}_{2^{a+b}}(q), 2^{x_1}y_1, \cdots, 2^{x_s}y_s)$,
where  ${\rm lcm}({\rm ord}_{2^{a+b}}(q), 2^{x_1}y_1, \cdots, 2^{x_s}y_s)$ denotes the least common multiple of the integers
${\rm ord}_{2^{a+b}}(q)$, $2^{x_1}y_1, \cdots, $ and $2^{x_s}y_s$.

Observe that $n=\frac{q^2+1}{10}$ is odd and ${\rm ord}_{rn}(q)=4$.
We have to consider two cases:

If $q\equiv1~(\bmod~4)$, then $a+b=1$; in this case, Theorem~\ref{dual-containing1} applies:
Suppose neither of $(i)$ and $(ii)$ satisfies.
We then know that $x_j>0$ for all $1\leq j\leq s$.  We are left to show that $(iii)$  holds. If $s=1$, it follows  from
$4=2^{x_1}y_1$ that $x_1=2$. This is impossible, since we would get that $(ii)$ holds. Assume that $x_1=x_2=\cdots=x_s=x=1$ ($s\geq2$);
this is a contradiction, because we would obtain $4={\rm lcm}(2y_1, \cdots, 2y_s)$.
Thus, condition $(iii)$ is satisfied.

If $q\equiv-1~(\bmod~4)$ then $a+b\geq2$, and so Theorem~\ref{dual-containing2} applies:
Suppose $(ii)$ and $(iii)$ are violated.
We then know that $a+b\leq e$, where $2^e\Vert(q+1)$. Thus, ${\rm ord}_{2^{a+b}}(q)=2$.
By $4={\rm lcm}(2, 2^{x_1}y_1, \cdots, 2^{x_s}y_s)$ again, we know that $(iv)$ satisfies.
\end{proof}

\section{New quantum MDS codes}
A $q$-ary quantum code $Q$ of
length $n$ and size $K$ is a $K$-dimensional subspace of the $q^n$-dimensional Hilbert space $(\mathbb{C}^q)^{\otimes n}$.
Let $k=\log_q(K)$. We use $[[n,k,d]]_q$ to denote a $q$-ary quantum code of length $n$ with size $q^k$ and minimum distance $d$.
Just as in the classical case,  the
parameters of an  $[[n,k,d]]_q$ quantum code must
satisfy the quantum Singleton bound: $2d\leq n-k+2$ (see \cite{Knill} and \cite{Ketkar}).
A quantum code
beating this quantum Singleton bound is called a quantum
maximum-distance-separable (MDS) code. Ketkar {\it et al.} in \cite{Ketkar} pointed out that if the classical MDS conjecture holds, then the length
of nontrivial quantum MDS codes cannot exceed $q^2+1$.

In this section, several classes of dual-containing MDS constacyclic codes are constructed and their parameters are computed.
Consequently, new  quantum MDS codes are derived from these parameters.

\subsection{New quantum MDS codes of length $\frac{q^2-1}{3}$}
In this subsection, we assume that
$q$ is an odd prime power with $3\mid(q+1)$.
Let $n=\frac{q^2-1}{3}$ and $r=3$. It is clear that $rn=q^2-1$, and hence every $q^2$-cyclotomic coset modulo $rn$ contains exactly one element.
Assume that $\eta\in \mathbb{F}_{q^2}$ is a primitive $3n$th root of unity, and let $\lambda=\eta^n$.

We will obtain  quantum MDS codes with parameters $[[\frac{q^2-1}{3},\frac{q^2-1}{3}-2d+2,d]]_q$,
where $2\leq d\leq\frac{2(q-2)}{3}+1$.
Using the Hermitian construction,  it suffices to construct  dual-containing  $\lambda$-constacyclic codes over $\mathbb{F}_{q^2}$
with parameters $[\frac{q^2-1}{3}, \frac{q^2-1}{3}-\frac{2(q-2)}{3},\frac{2(q-2)}{3}+1]$.
Note that these quantum MDS codes have minimum distance bigger than $\frac{q}{2}+1$ when $q>11$.

Corollary~\ref{exist1} guarantees  that dual-containing $\lambda$-constacyclic codes of length $n=\frac{q^2-1}{3}$
over $\mathbb{F}_{q^2}$ exist.
Let $C$ be a $\lambda$-constacyclic code with defining set
$$
Z_1=\{1+3(\frac{q-2}{3}+j)\,|\,1\leq j\leq\frac{2(q-2)}{3}\}.
$$
It is easy to see that $0<\frac{q-2}{3}+\frac{2(q-2)}{3}<n$, which implies that $|Z_1|=\frac{2(q-2)}{3}$ and that $C$ is an MDS $\lambda$-constacyclic code
with parameters $[n, n-\frac{2(q-2)}{3},\frac{2(q-2)}{3}+1]$. From the Hermitian construction, we only need to show that $Z_1\bigcap(-qZ_1)=\emptyset$.
\begin{lem}
With respect to the above notation, the $\lambda$-constacyclic code with defining set $Z_1$ over $\mathbb{F}_{q^2}$ is a dual-containing code.
\end{lem}
\begin{proof}
Suppose otherwise that $Z_1\bigcap(-qZ_1)\neq\emptyset$. Hence, two integers $j,k$ with $1\leq j,k\leq\frac{2(q-2)}{3}$ can be found such that
$$
-q\big(1+3(\frac{q-2}{3}+j)\big)\equiv1+3(\frac{q-2}{3}+k)~(\bmod~q^2-1).
$$
Expanding  this equation
gives,
\begin{equation}\label{equation}
3jq+3k\equiv0~(\bmod~q^2-1).
\end{equation}

If $1\leq j\leq \frac{q-2}{3}$, then $3jq+3k\leq3q\cdot\frac{q-2}{3}+3\cdot\frac{2(q-2)}{3}=q^2-4<q^2-1$, which contradicts
(\ref{equation}).

We are thus left to consider the case $\frac{q-2}{3}<j\leq \frac{2(q-2)}{3}$. Write $t=\frac{2(q-2)}{3}-j$, and so
$0\leq t<\frac{q-2}{3}$. Then $3jq+3k=3q(\frac{2(q-2)}{3}-t)+3k=2q^2-4q-3qt+3k$.
It follows from (\ref{equation}) that
$$
3qt+2q\equiv3k+2-2q~(\bmod~q^2-1).
$$
Observe that $0<3qt+2q<3q\cdot\frac{1}{3}(q-2)+2q=q^2$ and $5-2q\leq3k+2-2q\leq-2$. Thus,
$0<q^2-1+(3k+2-2q)<q^2-1$. At this point, two cases may occur: If $3qt+2q=q^2-1$, then
$q^2-1+(3k+2-2q)=0$ which is impossible; if $3qt+2q<q^2-1$, then
$3qt+2q=q^2-1+3k+2-2q$. Write $t'=\frac{1}{3}(q-2)-t$. Clearly, $t'>0$. We then have
$3q(\frac{1}{3}(q-2)-t')+4q=q^2+3k+1$, or equivalently, $2q-3qt'=3k+1$. This is a contradiction again, since
$2q-3qt'<0$ and $3k+1>0$.
\end{proof}

We have constructed a  dual-containing  $\lambda$-constacyclic code over $\mathbb{F}_{q^2}$
with parameters $[\frac{q^2-1}{3}, \frac{q^2-1}{3}-\frac{2(q-2)}{3},\frac{2(q-2)}{3}+1]$.
Since every $q^2$-cyclotomic coset contains precisely one element, we can obtain dual-containing
$\lambda$-constacyclic MDS codes with parameters $[\frac{q^2-1}{3}, \frac{q^2-1}{3}-\delta, \delta+1]$,
where $1\leq \delta\leq\frac{2(q-2)}{3}$.
Using the Hermitian construction, we have the following new quantum MDS codes.
\begin{Theorem}\label{three}
Let $q$ be an odd prime power with $3\mid (q+1)$. Then, there exist quantum MDS codes with parameters $[[\frac{q^2-1}{3},\frac{q^2-1}{3}-2d+2,d]]_q$,
where $2\leq d\leq\frac{2(q-2)}{3}+1$.
\end{Theorem}

\begin{Example}\label{3example1}
Take $q=11$, and so $n=40$. Using Theorem~\ref{three}, we get six quantum MDS codes with parameters
$[[40,28,7]]_{11},[[40,30,6]]_{11}, [[40,32,5]]_{11}, [[40,34,4]]_{11}, [[40,36,3]]_{11}, [[40,38,2]]_{11}$.
\end{Example}

\begin{Example}
Take $q=17$, and so $n=96$. Using Theorem~\ref{three}, we get ten quantum MDS codes.
We list five of them:
$[[96,76,11]]_{17},[[96,78,10]]_{17}, [[96,80,9]]_{17}, [[96,82,8]]_{17}, [[96,84,7]]_{17}$.
\end{Example}

\begin{Example}\label{3example3}
Let  $q=23$, and so $n=176$. Using Theorem~\ref{three} produces $14$ new quantum MDS codes.
We list five of them: $[[176,148,15]]_{23},[[176,150,14]]_{23}, [[176,152,13]]_{23}, [[176,154,12]]_{23}, [[176,156,11]]_{23}$.

\end{Example}
\subsection{New quantum MDS codes of length $\frac{q^2-1}{5}$}
Let
$q$ be an odd prime power with $5\mid(q+1)$.
Let $n=\frac{q^2-1}{5}$ and $r=5$. It is readily seen that every $q^2$-cyclotomic coset modulo $rn$ contains exactly one element.
Assume that $\eta\in \mathbb{F}_{q^2}$ is a primitive $5n$th root of unity, and let $\lambda=\eta^n$.

In this subsection, we will obtain  quantum MDS codes with parameters $[[\frac{q^2-1}{5},\frac{q^2-1}{5}-2d+2,d]]_q$,
where $2\leq d\leq\frac{3(q+1)}{5}-1$. Note  that these quantum MDS codes
with minimum distance bigger than $\frac{q}{2}+1$ once $q>19$.

As we did previously, we only need to construct  dual-containing  $\lambda$-constacyclic  MDS codes over $\mathbb{F}_{q^2}$
with parameters $[\frac{q^2-1}{5}, \frac{q^2-1}{5}-\frac{3(q+1)}{5}+2,\frac{3(q+1)}{5}-1]$.
Note also that Corollary~\ref{exist1} guarantees  that dual-containing $\lambda$-constacyclic codes of length $n=\frac{q^2-1}{5}$
over $\mathbb{F}_{q^2}$ exist.

Let $C$ be a $\lambda$-constacyclic code with defining set
$$
Z_2=\{1+5(\frac{2(q+1)}{5}-1+j)\,|\,1\leq j\leq\frac{3(q+1)}{5}-2\}.
$$
It is easy to see that $0<\frac{2(q+1)}{5}-1+\frac{3(q+1)}{5}-2<n$, which implies that $|Z_2|=\frac{3(q+1)}{5}-2$ and that $C$ is an MDS $\lambda$-constacyclic code
with parameters $[\frac{q^2-1}{5}, \frac{q^2-1}{5}-\frac{3(q+1)}{5}+2,\frac{3(q+1)}{5}-1]$.
We are left to show that $Z_2\bigcap(-qZ_2)=\emptyset$.
\begin{lem}\label{similar}
With respect to the above notation, the $\lambda$-constacyclic code with defining set $Z_2$ over $\mathbb{F}_{q^2}$ is a dual-containing code.
\end{lem}
\begin{proof}
Suppose otherwise that $Z_2\bigcap(-qZ_2)\neq\emptyset$. Then, we can find two integers $j,k$ with $1\leq j,k\leq\frac{3(q+1)}{5}-2$  such that
$$
-q\big(1+5(\frac{2(q+1)}{5}-1+j)\big)\equiv1+5(\frac{2(q+1)}{5}-1+k)~(\bmod~q^2-1).
$$
We then have
\begin{equation}\label{equation2}
5jq+5k\equiv0~(\bmod~q^2-1).
\end{equation}
We analyze (\ref{equation2}) in a series of cases.

If $1\leq j\leq\frac{q+1}{5}-1$, then $5q\leq 5jq\leq q^2-4q$ and $5\leq5k\leq3q-7$. It follows that
$5jq+5k\leq q^2-q-7<q^2-1$, and hence $5jq+5k=0$, which is impossible.

If $j=\frac{q+1}{5}$, then routine computations show that $q^2-1<5jq<2(q^2-1)$. It follows from
(\ref{equation2}) that $5jq-(q^2-1)=q^2-1-5k$, or equivalently $q+5k+2=q^2$. Note that $5k\leq3q-7$.
Thus, $q+5k+2\leq4q-5<q^2$,
which is a contradiction.

If $\frac{q+1}{5}+1\leq j\leq\frac{2(q+1)}{5}-1$, then $6q+1\leq 5jq-(q^2-1)\leq q^2-3q+1$.
By (\ref{equation2}) again, we have $5jq-(q^2-1)=q^2-1-5k$. That is, $5jq+5k=2(q^2-1)$.
But we see that $5jq+5k\leq2q^2-7$, which is a contradiction.

If $\frac{2(q+1)}{5}\leq j\leq\frac{3(q+1)}{5}-2$, then $2q+2\leq 5jq-2(q^2-1)\leq q^2-7q+2$.
It follows from (\ref{equation2}) that $5jq-2(q^2-1)=q^2-1-5k$, or equivalently,
$5jq+5k=3(q^2-1)$.  Now $5jq+5k\leq3q^2-4q-7<3(q^2-1)$ gives the desired contradiction.
\end{proof}

Since every $q^2$-cyclotomic coset contains precisely one element, we can obtain dual-containing
$\lambda$-constacyclic MDS codes with parameters $[\frac{q^2-1}{5}, \frac{q^2-1}{5}-\delta,\delta+1]$,
where $1\leq \delta\leq\frac{3(q+1)}{5}-2$.
Using the Hermitian construction, we have the following new quantum MDS codes.

\begin{Theorem}\label{five}
Let $q$ be an odd prime power with $5\mid (q+1)$. Then, there exist quantum MDS codes with parameters $[[\frac{q^2-1}{5},\frac{q^2-1}{5}-2d+2,d]]_q$,
where $2\leq d\leq\frac{3(q+1)}{5}-1$.
\end{Theorem}

\begin{Example}
Let  $q=9$, and so $n=16$. Using Theorem~\ref{five} produces $4$  quantum MDS codes with parameters
$[[16,8,5]]_{9},[[16,10,4]]_{9}, [[16,12,3]]_{9}, [[16,14,2]]_{9}$.
\end{Example}

\begin{Example}
Let  $q=19$, and so $n=72$. Using Theorem~\ref{five} produces $10$  quantum MDS codes.
We list five of them: $[[72,52,11]]_{19},[[72,54,10]]_{19}, [[72,56,9]]_{19}, [[72,58,8]]_{19}, [[72,60,7]]_{19}$.
\end{Example}

\begin{Example}
Let  $q=29$, and so $n=168$. Using Theorem~\ref{five} produces $16$  quantum MDS codes.
We list five of them: $[[168,136,17]]_{29},[[168,138,16]]_{29}, [[168,140,15]]_{29}, [[168,142,14]]_{29}, [[168,144,13]]_{29}$.
\end{Example}

\subsection{New quantum MDS codes of length $\frac{q^2-1}{7}$}
Let
$q$ be an odd prime power with $7\mid(q+1)$.
Let $n=\frac{q^2-1}{7}$ and $r=7$. It is readily seen that every $q^2$-cyclotomic coset modulo $rn$ contains exactly one element.
Assume that $\eta\in \mathbb{F}_{q^2}$ is a primitive $7n$th root of unity, and let $\lambda=\eta^n$.

In this subsection, we will obtain  quantum MDS codes with parameters $[[\frac{q^2-1}{7},\frac{q^2-1}{7}-2d+2,d]]_q$,
where $2\leq d\leq\frac{4(q+1)}{7}-1$. Note  that these quantum MDS codes
with minimum distance bigger than $\frac{q}{2}+1$ once $q>27$.

Let $C$ be a $\lambda$-constacyclic code with defining set
$$
Z_3=\{1+5(\frac{3(q+1)}{7}-1+j)\,|\,1\leq j\leq\frac{4(q+1)}{7}-2\}.
$$
It is easy to see that $0<\frac{3(q+1)}{7}-1+\frac{4(q+1)}{7}-2<n$, which implies that $|Z_3|=\frac{4(q+1)}{7}-2$ and that $C$ is an MDS $\lambda$-constacyclic code
with parameters $[\frac{q^2-1}{7}, \frac{q^2-1}{7}-\frac{4(q+1)}{7}+2,\frac{4(q+1)}{7}-1]$.
To achieve our goal, we are left to show that $Z_3\bigcap(-qZ_3)=\emptyset$.
Since the proof is very similar to Lemma~\ref{similar},  we just state the fact and omit its proof.
\begin{lem}
With respect to the above notation, the $\lambda$-constacyclic code with defining set $Z_3$ over $\mathbb{F}_{q^2}$ is a dual-containing code.
\end{lem}

\begin{Theorem}\label{seven}
Let $q$ be an odd prime power with $7\mid (q+1)$. Then, there exist quantum MDS codes with parameters $[[\frac{q^2-1}{7},\frac{q^2-1}{7}-2d+2,d]]_q$,
where $2\leq d\leq\frac{4(q+1)}{7}-1$.
\end{Theorem}

\begin{Example}
Let  $q=13$, and so $n=24$. Using Theorem~\ref{seven} produces $6$  quantum MDS codes with parameters
$[[24,12,7]]_{13},[[24,14,6]]_{13}, [[24,16,5]]_{13}, [[24,18,4]]_{13}, [[24,20,13]]_{13}, [[24,22,13]]_{13}$.
\end{Example}

\begin{Example}
Take $q=27$, and so $n=104$.  Using Theorem~\ref{seven} derives  $14$  quantum MDS codes.
We list five of them: $[[104,76,15]]_{27},[[104,78,14]]_{27}, [[104,80,13]]_{27}, [[104,82,12]]_{27}, [[104,84,11]]_{27}$.
\end{Example}

\subsection{New quantum MDS codes of length $\frac{q^2+1}{10}$}
Let $q$ be an odd prime power such that $10\mid(q^2+1)$, i.e., $q$ has  the form $10m+3$ or $10m+7$, where $m$ is a positive integer.
Let $n=\frac{q^2+1}{10}$ and $r=q+1$.
Assume that $\eta\in \mathbb{F}_{q^2}^*$ is a primitive $rn$th root of unity, and let $\lambda=\eta^n$.

Note that Corollary~\ref{exist2} guarantees  that dual-containing $\lambda$-constacyclic codes of length $n=\frac{q^2+1}{10}$
over $\mathbb{F}_{q^2}$ exist.
To construct quantum MDS codes,
we first construct a family of MDS dual-containing constacyclic  codes over $\mathbb{F}_{q^2}$ of length $\frac{q^2+1}{10}$, as we show below.

\begin{lem}\label{cyclotomic}
Let $n=\frac{q^2+1}{10}$, $s=\frac{q^2+1}{2}$ and $r=q+1$. Then $\theta_{r,n}=\{1+ri\,|\,0\leq i\leq n-1\}$ is a disjoint union of $q^2$-cyclotomic cosets:
$$
\theta_{r,n}=C_{s}\bigcup\big(\bigcup\limits_{k=0}^{\frac{n-1}{2}-1}C_{s-(q+1)(\frac{n-1}{2}-k)}\big).
$$
\end{lem}
\begin{proof}
Observe that $q^4\equiv1~(\bmod~rn)$ and $q^2\not\equiv1~(\bmod~rn)$, which imply that each $q^2$-cyclotomic coset contains one or two elements.
Now,
$$
q^2\big(1+(q+1)j\big)=q^2+q^2(q+1)j=q^2+(q^2+1-1)(q+1)j\equiv q^2-(q+1)j~(\bmod~rn).
$$
It is clear that for $0\leq j\leq n-1$, $1+(q+1)j\equiv q^2-(q+1)j~(\bmod~rn)$ if and only if $j=\frac{q-1}{2}+tn$ ($t$ is an integer),
which forces $t=0$ and hence $j=\frac{q-1}{2}$. This shows that $s=1+(q+1)\frac{q-1}{2}=\frac{q^2+1}{2}$ is the unique element of $\theta_{r,n}$ with
$q^2s\equiv s~(\bmod~rn)$.
To complete the proof, it suffices to show that for any $0\leq i\neq j\leq\frac{n-1}{2}-1$,
$C_{s-(q+1)(\frac{n-1}{2}-i)}=\{s-(q+1)(\frac{n-1}{2}-i),s+(q+1)(\frac{n-1}{2}-i)\}$
and $C_{s-(q+1)(\frac{n-1}{2}-j)}$ are distinct.
Suppose otherwise that $C_{s-(q+1)(\frac{n-1}{2}-i)}=C_{s-(q+1)(\frac{n-1}{2}-j)}$ for some $0\leq i\neq j\leq\frac{n-1}{2}-1$.
If $s-(q+1)(\frac{n-1}{2}-i)\equiv s-(q+1)(\frac{n-1}{2}-j)~(\bmod~rn)$,  then $i\equiv j~(\bmod~n)$ which is impossible;
If  $s-(q+1)(\frac{n-1}{2}-i)\equiv s+(q+1)(\frac{n-1}{2}-j)~(\bmod~rn)$, then $i+j\equiv -1~(\bmod~n)$ which is a contradiction.
\end{proof}

Let $C$ be a $\lambda$-constacyclic code of length $n=\frac{q^2+1}{10}$ over $\mathbb{F}_{q^2}$
with defining set
$$
Z_4=\bigcup\limits_{j=0}^{2m-1}C_{s-(q+1)(\frac{n-1}{2}-j)}.
$$
We then know from Lemma~\ref{cyclotomic} that $Z_4$ is a disjoint union of $q^2$-cyclotomic cosets modulo $rn$ with $|Z_4|=4m$.
Moreover, we assert that the minimum distance of $C$ is equal to $4m+1$. To see this, observe that
$$
Z_4=\big\{s+r(\frac{n-1}{2}-2m+1), s+r(\frac{n-1}{2}-2m+2),\cdots, s+r(\frac{n-1}{2}-1),$$ $$
s+r\frac{n-1}{2}, s-r\frac{n-1}{2}, s-r(\frac{n-1}{2}-1),\cdots, s-r(\frac{n-1}{2}-2m+1)\big\}.
$$
A simple calculation shows that $s+r\frac{n-1}{2}+r\equiv s-r\frac{n-1}{2}~(\bmod~rn)$.
By the BCH bound for constacyclic  codes, $C$ is an MDS code with parameters $[n, n-4m, 4m+1]$.

We will prove that $C$ is a dual-containing code, so that an MDS quantum code with parameters $[[n, n-8m, 4m+1]]_q$ can be constructed.
\begin{lem}
Let $Z_4$ be defined as above. Let $C$ be a $\lambda$-constacyclic code of length $n=\frac{q^2+1}{10}$ over $\mathbb{F}_{q^2}$
with defining set $Z_4$. Then $C$ is a dual-containing code.
\end{lem}
\begin{proof}
We have to prove that $Z_4\bigcap(-qZ_4)=\emptyset$. We just give a proof for the case $q=10m+3$. The  case for $q=10m+7$ is proved just as similarly.
Suppose there exist integers $j,k$ with $0\leq j,k\leq 2m-1$ such that $C_{-q(s-(q+1)(\frac{n-1}{2}-j))}=C_{s-(q+1)(\frac{n-1}{2}-k)}$.
Write $j=j_1m+j_0$ and $k=k_1m+k_0$, where $j_1,k_1\in\{0,1\}$ and $0\leq j_0,k_0<m$. Let  $j_0'=m-j_0$ and $k_0'=m-k_0$, and so $0<j_0', k_0'\leq m$.

{\it Case I.}~~$-q(s-(q+1)(\frac{n-1}{2}-j))\equiv s-(q+1)(\frac{n-1}{2}-k)~(\bmod~rn)$.
After routine computations, we obtain
\begin{equation}\label{one}
-\frac{q+1}{2}\equiv qj+k~(\bmod~n).
\end{equation}
Now $qj+k=(10m+3)(j_1m+j_0)+k_1m+k_0=10j_1m^2+(10j_0+3j_1+k_1)m+3j_0+k_0=10j_1m^2+(10m-10j_0'+3j_1+k_1)m+3m-3j_0'+m-k_0'$.

 Assume
$qj+k<n$.

If $j_1=0$, it follows from (\ref{one}) that
$$
(10m-10j_0'+k_1)m+3m-3j_0'+m-k_0'=n-\frac{q+1}{2}=10m^2+m-1.
$$
This leads to
$$
(k_1-10j_0'+4)m=m+3j_0'+k_0'-1,
$$
which is a contradiction,  since $(k_1-10j_0'+4)m<0$ and $m+3j_0'+k_0'-1>0$.

If $j_1=1$, then
$$
10m^2+(10m-10j_0'+3+k_1)m+3m-3j_0'+m-k_0'=10m^2+m-1,
$$
or equivalently,
$
10j_0'm+k_0'=1+(k_1+10m+6)m-3j_0'.
$
Now, $10j_0'm+k_0'\leq10m^2+m$,  but $1+(k_1+10m+6)m-3j_0'>10m^2+3m$, which is a contradiction.

Assume
$qj+k>n$.

If $j_1=0$, then $qj+k=(10m+3)j_0+k<(10m+3)m+2m=10m^2+5m<n$; this is impossible.

If $j_1=1$, we claim that $qj+k-n=(k_1+10j_0-3)m+k_0+3j_0-1<n$; this is because
$(k_1+10j_0-3)m+k_0+3j_0-1\leq(1+10m-10-3)m+m-1+3(m-1)-1<n$. From (\ref{one}) again,
we have $(k_1+10j_0-3)m+k_0+3j_0-1=n-\frac{q+1}{2}=10m^2+m-1$, or equivalently,
$(k_1-10j_0')m=k_0'+3j_0'$. This is a contradiction, because $k_0'>0, j_0'>0$
and $k_1-10j_0'<0$.

{\it Case II.}~~$-q(s-(q+1)(\frac{n-1}{2}-j))\equiv s+(q+1)(\frac{n-1}{2}-k)~(\bmod~rn)$.
After routine computations, we get
\begin{equation}\label{two}
-\frac{q-1}{2}\equiv qj-k~(\bmod~n).
\end{equation}
As we did previously, $qj-k=(10m+3)(j_1m+j_0)-k_1m-k_0=10j_1m^2+(10j_0+3j_1-k_1)m+3j_0-k_0$.

If $j_1=0$, then $qj-k\leq(10m+3)(m-1)<n$. When $0<qj-k<n$,  by (\ref{two}), $10j_0m+3j_0-k_1m-k_0=10m^2+m$,
which is equivalent to $10j_0'm-m+3j_0'+k_1m-k_0'=0$. This is impossible, since
$10j_0'm-m+3j_0'+k_1m-k_0'>10m-m-m>0.$
When $qj-k<0$ (Clearly, $0<k-qj<n$), we obtain $5m+1=\frac{q-1}{2}=k-qj$, which is a contradiction since $k<2m$.

If $j_1=1$ and $j_0=0$, we have  $qj-k=(10m+3)m-k<n$.
Using (\ref{two}), we get $k=2m$, also a contradiction.

If $j_1=1$ and $j_0>0$, we then know that $qj-k=10m^2+(10j_0+3-k_1)m+3j_0-k_0>n$.
On the other hand, $qj-k-n=10m^2-10j_0'm-3j_0'-k_1m-m+k_0'-1<n.$
Applying (\ref{two}) again, we obtain $-10j_0'm-3j_0'-k_1m-2m+k_0'-1=0$.
This is impossible, because $-10j_0'm-3j_0'-k_1m-2m+k_0'-1<0$.
\end{proof}

Since every $q^2$-cyclotomic coset  contained in $Z_4$ has two elements, we can obtain dual-containing
$\lambda$-constacyclic MDS codes with parameters $[\frac{q^2+1}{10}, \frac{q^2+1}{10}-\delta,\delta+1]$,
where $2\leq \delta\leq 4m$ is even.
Using the Hermitian construction, we have the following new quantum MDS codes.

\begin{Theorem}\label{last}
Let $q$ be an odd prime power having the form $q=10m+3$ or $q=10m+7$. Then, there exist quantum MDS codes with parameters $[[\frac{q^2+1}{10},\frac{q^2+1}{10}-2d+2,d]]_q$,
where $3\leq d\leq4m+1$ is odd.
\end{Theorem}

\begin{Example}
Take $q=13$, and so $n=17$.  Using Theorem~\ref{last} derives  $2$ new quantum MDS codes.
We list five of them: $[[17,9,5]]_{13}, [[17,13,3]]_{13}$.
\end{Example}

\begin{Example}
Take $q=17$, and so $n=29$.  Using Theorem~\ref{last} derives  $2$ new quantum MDS codes.
We list five of them: $[[29,21,5]]_{17},[[29,25,3]]_{17}$.
\end{Example}

\begin{Example}
Take $q=23$, and so $n=53$.  Using Theorem~\ref{last} derives  $4$ new quantum MDS codes
 with parameters $[[53,37,9]]_{23}, [[53,41,7]]_{23}, [[53,45,5]]_{23}, [[53,49,3]]_{23}$.
\end{Example}

\section{Summary and code comparisons}
In this section we first list the parameters of all known quantum MDS codes. Then we compare the parameters of quantum MDS codes available in the literature
with the parameters of the new quantum codes.
\begin{center}
\begin{longtable}{c|c|c|c}  
\caption{Quantum MDS codes}\\\hline
$Class$  & Length  &  Distance  &  Reference     \\\hline
$1$  &  $n\leq q+1$  &  $d\leq\lfloor n/2\rfloor+1$  &  \cite{Feng02}, \cite{Grassl}, \cite{Grassl2} \\
\hline
$2$  &  $mq-l$  &  $d\leq m-l+1$,  &  \cite{Li}, \cite{SK1}  \\
    &          &  $0\leq l< m, 1<m<q$ &    \\
\hline
$3$  &  $mq-l$  &  $3\leq d\leq (q+1-\lfloor l/m\rfloor)/2$,  &  \cite{Jin10}  \\
    &          &  $0\leq l \leq q-1, 1\leq m\leq q$   &    \\
\hline
$4$  &  $r(q-1)+1$  &  $d\leq (q+r+1)/2$  &  \cite{Jin14} \\
&  $q\equiv r-1 \,\,{\rm mod} \,\,2r$  &       &    \\
\hline
$5$  &  $q^2-s$  &  $q/2+1<d\leq q-s$  &  \cite{Jin14} \\
&  $0\leq s<q/2-1$  &       &    \\
\hline
$6$  &  $(q^2+1)/2-s$  &  $q/2+1<d\leq q-s$  &  \cite{Jin14} \\
&  $0\leq s<q/2-1$  &       &    \\
\hline
$7$  &  $(q^2+1)/2$, $q$ \,odd   &  $3\leq d \leq q$, $d$ \,odd  &  \cite{Kai13}\\
\hline
$8$  &  $4\leq n \leq q^2+1$  &  $3$  &  \cite{Bierbrauer00},\cite{Jin10}, \cite{LX}  \\
&  $q\neq 2 \,\,and  \,\,n\neq 4$  &       &    \\
\hline
$9$  &  $q^2-l$  &  $ d \leq q-l, 0\leq l \leq q-2$  &  \cite{Grassl},\cite{Li}\\
\hline
$10$  &  $q^2+1$  &  $2\leq d \leq q+1$  &  \cite{Jin10}, \cite{Jin14}, \cite{Kai13}, \cite{Guardia11}\\
\hline
$11$ &  $(q^2-1)/2$, $q$ \,odd & $2\leq d \leq q$ &  \cite{Kai14}\\
\hline
$12$ & $\frac{q^2-1}{r}$, $q$ odd  &  $2\leq d \leq (q+1)/2$  &  \cite{Kai14}\\
     & $r\mid(q+1)$,  $r$  even and $r\neq 2$  &    &   \\
\hline
$13$   &   $\lambda(q+1)$, $q$ odd  &  $2\leq d \leq (q+1)/2+\lambda$  & \cite{Kai14}\\
       &   $\lambda$ odd, \, $\lambda\mid(q-1)$  &    &    \\
\hline
$14$   &   $2\lambda(q+1)$, $q\equiv 1\,{\rm mod }\,4$  &  $2\leq d \leq (q+1)/2+2\lambda$  & \cite{Kai14}\\
       &   $\lambda$ odd, $\lambda|(q-1)$  &    &    \\
\hline
$15$   &   $(q^2+1)/5$, $q=20m+3$  &   $2 \leq d \leq (q+5)/2$, &   \cite{Kai14}\\
       &   or \, $20m+7$  &   $d$ even &   \\
\hline
$16$   &   $(q^2+1)/5$, $q=20m-3$  &   $2 \leq d \leq (q+3)/2$, &   \cite{Kai14}\\
       &   or \, $20m-7$  &  $d$ even &   \\
\hline
$17$   &   $n=\frac{q^2-1}{3}$, $3\mid(q+1)$  &  $2\leq d\leq\frac{2(q-2)}{3}+1$, &   {\bf New}\\
          & &   \\
\hline
$18$   &    $n=\frac{q^2-1}{5}$, $5\mid(q+1)$  &    $2\leq d\leq\frac{3(q+1)}{5}-1$, &   {\bf New}\\
      &  &   \\
\hline
$19$   &   $n=\frac{q^2-1}{7}$, $7\mid(q+1)$  &  $2\leq d\leq\frac{4(q+1)}{7}-1$, &   {\bf New}\\
       & &   \\
\hline
$20$   &  $n=\frac{q^2+1}{10}$  &   $3\leq d\leq4m+1$, &   {\bf New}\\
       &   $q=10m+3$ or $q=10m+7$  &  $d$ odd &   \\
\hline

\end{longtable}
 \end{center}

We have constructed  four classes of $q$-ary
quantum MDS codes (see Classes 17-20 in Table 1).
Fixing an odd prime power $p$ and comparing  the lengths in Table $1$, we find that only Class $3$ has   the possibility to reach these lengths.
The following examples show that the new codes have minimum distance
greater than the ones available in the literature.

\begin{center}
\begin{longtable}{c|c|c|c}  
\caption{Code comparisons}\\\hline
$q$  & Length  &  $d$ (Class $17$) & $d$ (Class $3$)     \\\hline

$11$  &  $40$  &  $7$  &  $5$  \\
\hline
$17$  &  $96$  &  $11$  &  $8$ \\

\hline
$23$  &  $176$  &  $15$  &  $11$  \\

\hline

\end{longtable}
 \end{center}

 \begin{center}
\begin{longtable}{c|c|c|c}  
\caption{Code comparisons}\\\hline
$q$  & Length  &  $d$ (Class $18$) & $d$ (Class $3$)     \\\hline
$9$  &  $16$  &  $5$  &  $4$ \\
\hline
$19$  &  $72$  &  $11$  &  $9$  \\

\hline
$29$  &  $168$  &  $17$  &  $14$  \\

\hline

\end{longtable}
 \end{center}

  \begin{center}
\begin{longtable}{c|c|c|c}  
\caption{Code comparisons}\\\hline
$q$  & Length  &  $d$ (Class $19$) & $d$ (Class $3$)     \\\hline
$13$  &  $24$  &  $7$  &  $6$ \\
\hline
$27$  &  $104$  &  $15$  &  $13$  \\
\hline

\end{longtable}
 \end{center}

 \begin{center}
\begin{longtable}{c|c|c|c}  
\caption{Code comparisons}\\\hline
$q$  & Length  &  $d$ (Class $20$) & $d$ (Class $3$)     \\\hline
$13$  &  $17$  &  $5$  &  $5$ \\
\hline

$23$  &  $53$  &  $9$  &  $8$  \\

\hline

\end{longtable}
 \end{center}

\end{document}